\documentclass[aps,pra,preprint,showpacs]{revtex4}
\usepackage{amsmath}
\usepackage{amsfonts}
\usepackage{amssymb}
\usepackage{dcolumn}

\newcolumntype{w}[1]{D{.}{.}{#1}}
\newcolumntype{.}{D{x}{}{-1}}

\begin{document}
\preprint{Version 1.0}

\title{Efficient approach to two-centre exponential integrals with applications
       to excited states of molecular hydrogen}

\author{Krzysztof Pachucki}
\affiliation{Faculty of Physics, University of Warsaw,
             Ho\.{z}a 69, 00-681 Warsaw, Poland}

\begin{abstract}
A general approach to evaluation of two-centre two-electron exponential integrals
with arbitrary parameters is presented. The results for the Born-Oppenheimer
potential for various excited states of molecular hydrogen with Ko\l os-Wolniewicz
functions are obtained with precision exceeding previous values by about 3 orders of magnitude.
\end{abstract}
\pacs{31.15.ac, 31.15.-p, 31.15.vn}
\maketitle

\section{Introduction}
The hydrogen molecule was a test of quantum mechanics since its early beginnings.
The calculations of the H$_2$ dissociation energy by Ko\l os and Wolniewicz \cite{kw1, kw2}
were more accurate than experiments at that time \cite{heg1} and led to the verification of experimental values \cite{heg2, stw}.
At present theoretical predictions for the dissociation energy \cite{komasa:11} are in perfect agreement
with the most recent experiments \cite{liu:09, liu:10, dickenson:13}
and test the validity of quantum electrodynamic theory (QED) in molecular systems.
Moreover, assuming that QED  theory is correct, the comparison with experimental spectra puts
strong bounds on the unknown long-range interaction between hadrons \cite{salumbides}.
In this work we intend to extend the high-precision results obtained for the ground electronic
state of H$_2$ to excited states, where accurate measurements have recently became feasible \cite{salumb_private}.
The principal problem is the accurate solution of the Schr\"odinger equation. It is surprising
that the calculations by Wolniewicz {\em et al.} years ago \cite{sigma, OSW99}
have been surpassed only for the lowest lying states of H$_2$ \cite{RCK94, CKR95, KC03, RK_book}
by calculations based on explicitly correlated Gaussian functions (ECG).

The aim of this work is to present a computational approach to
the nonrelativistic energies of the H$_2$ molecule using an explicitly correlated basis of the form
\begin{align}
\phi =  &e^{- y\,(r_{1A}-r_{1B}) - x\, (r_{2A}-r_{2B}) - u\,(r_{1A}+r_{1B}) - w\, (r_{2A}+r_{2B})} \nonumber \\&\times
  r_{12}^{n_0}\,(r_{1A}-r_{1B})^{n_1}\,(r_{2A} - r_{2B})^{n_2}\,(r_{1A}+ r_{1B})^{n_3}\,(r_{2A}+r_{2B})^{n_4}\,,
\label{01}
\end{align}
where $u, w, x$ and $y$ are real nonlinear parameters,
the subscripts $1$ and $2$ numerate the electrons, and $A$ and $B$ numerate the nuclei.
Thus, $r_{12}$ is the interelectronic distance, whereas
$r_{1A}$ is the distance between the first electron and the nuclei $A$.
This basis was introduced by Ko\l os and Wolniewicz \cite{kw1} to obtain the first accurate
results for ground and excited states of two-electron diatomic molecules.
Integrals with these functions were performed by using the Neumann expansion of $1/r_{12}$
in spherical oblate coordinates, see the most recent review by Harris \cite{harris2}.
In the approach developed here, these integrals are calculated by the Taylor expansion in $r = r_{AB}$,
the internuclear distance. Since this expansion is absolutely convergent for all
positive values of $r$, and all terms of expansion are simple algebraic and
logarithmic functions of nonlinear parameters,
with the help of a multiprecision floating-point library \cite{mpfun} one can obtain all the integrals
in Eq. (\ref{03}) with arbitrary powers of the electron distances.

\section{Master integral}
According to this approach, one considers in the first step
the so-called master two-electron two-centre integral $f(r)$,
\begin{equation}
f(r) = r\,\int \frac{d^3 r_1}{4\,\pi}\,\int \frac{d^3 r_2}{4\,\pi}\,
\frac{e^{-w_1\,r_{12} - u\,(r_{1A} + r_{1B}) - w\,(r_{2A} + r_{2B})
                    - y\,(r_{1A} - r_{1B}) - x\,(r_{2A} - r_{2B})}}
{r_{12}\,r_{1A}\,r_{1B}\,r_{2A}\,r_{2B}}. \label{02}
\end{equation}
Any additional power of electronic distances in the integral of Eq. (\ref{02})
can be obtained from $f(r)$ by differentiation with respect to the corresponding nonlinear parameter,
\begin{align}
&f(r,n_0,n_1,n_2,n_3,n_4) = \frac{1}{n_0!\,n_1!\,n_2!\,n_3!\,n_4!}\nonumber \\
&\times
\biggl(-\frac{\partial}{\partial w_1}\biggr)^{n_0}\biggr|_{w_1 = 0}\,
\biggl(-\frac{\partial}{\partial y}\biggr)^{n_1}\,
\biggl(-\frac{\partial}{\partial x}\biggr)^{n_2}\,
\biggl(-\frac{\partial}{\partial u}\biggr)^{n_3}\,
\biggl(-\frac{\partial}{\partial w}\biggr)^{n_4}\,f(r)\,,
\label{03}
\end{align}
where we include additional factorials in the denominator for simplicity of
recursion relations in the next section.
The master integral $f(r)$ is not known in a closed analytical form, but
it can be expressed in terms of a one-dimensional integral representation,
see Refs. \cite{rec_h2, rec2_h2} for details. Moreover, it can be shown \cite{rec2_h2, lesiuk:12}
that $f(r)$ satisfies the following differential equation
\begin{equation}
\biggl[\sigma_4\,\frac{d^2}{d\,r^2}\,r\,\frac{d^2}{d\,r^2} + \sigma_2\,\frac{d}{d\,r}\,r\,\frac{d}{d\,r}
+ \sigma_0\,r\biggr]\,f(r)  = F(r), \label{04}
\end{equation}
where
\begin{eqnarray}
\sigma &=&\sigma_0 +t^2\,\sigma_2 + t^4\,\sigma_4, \label{05}\\
\sigma_0 &=& w_1^2\,(u + w - x - y)\,(u - w + x - y)\,(u - w - x + y)\,(u + w + x + y) \nonumber \\&&
        + 16\,(w\,x - u\,y)\,(u\,x - w\,y)\,(u\,w - x\,y)\nonumber \\
         &=& \sigma_{00} + w_1^2\,\sigma_{02}\,,\nonumber \\
\sigma_2 &=& w_1^4-2\,w_1^2\,(u^2+w^2+x^2+y^2)+16\,u\,w\,x\,y\nonumber \\
         &=& \sigma_{20} + w_1^2\,\sigma_{22} + w_1^4\,, \nonumber \\
\sigma_4 &=& w_1^2\,,\nonumber
\end{eqnarray}
with the inhomogeneous term given by
\begin{eqnarray}
F(r) &=&
w_1\,\biggl(\frac{1}{r^2} + \frac{2\,w_1 + u + w + x - y}{r}\biggr)\,e^{-r\,(u + w + w_1 + x - y)}
\nonumber \\ && +
  w_1\,\biggl(\frac{1}{r^2} + \frac{2\,w_1 + u + w - x + y}{r}\biggr)\,e^{-r\,(u + w + w_1 - x + y)}
\nonumber \\ &&
- w_1\,\biggl(\frac{1}{r^2} + \frac{u + w - x - y}{r}\biggr)\,e^{-r\,(u + w - x - y)}
\nonumber \\ &&
- w_1\,\biggl(\frac{1}{r^2} + \frac{u + w + x + y}{r}\biggr)\,e^{-r\,(u + w + x + y)}
\nonumber \\ &&
+ \biggl[\frac{w_1^2}{2}\,(u - w + x - y) + 2\,u\,w\,(y-x) + 2\,x\,y\,(w-u)\biggr]\,F_1
\nonumber \\ &&
+ \biggl[\frac{w_1^2}{2}\,(u - w - x + y) + 2\,u\,w\,(x-y) + 2\,x\,y\,(w-u)\biggr]\,F_2
\nonumber \\ &&
+ \biggl[\frac{w_1^2}{2}\,(u + w + x + y) + 2\,u\,w\,(x+y) + 2\,x\,y\,(u+w)\biggr]\,F_3
\nonumber \\ &&
+ \biggl[\frac{w_1^2}{2}\,(u + w - x - y) - 2\,u\,w\,(x+y) + 2\,x\,y\,(u+w)\biggr]\,F_4\,,
\label{06}
\end{eqnarray}
where
\begin{eqnarray}
         F_1 &=& {\rm Ei}[-r\,(w_1 + 2\,u)]\,\exp[r\,(u - w + x - y)]
                -{\rm Ei}[-r\,(w_1 + 2\,w)]\,\exp[-r\,(u - w + x - y)]\,,\nonumber \\[2ex]
         F_2 &=& {\rm Ei}[-r\,(w_1 + 2\,u)]\,\exp[r\,(u - w - x + y)]
                -{\rm Ei}[-r\,(w_1 + 2\,w)]\,\exp[-r\,(u - w - x + y)]\,,\nonumber \\ [2ex]
         F_3 &=& {\rm Ei}[-2\,r\,(u + w)]\,\exp[r\,(u + w + x + y)]
                +\biggl\{{\rm Ei}[2\,r\,(x + y)]-{\rm Ei}[-r\,(w_1 - 2\,x)]\nonumber \\
             & &-{\rm Ei}[-r\,(w_1 - 2\,y)]
                -\ln\biggl[\frac{(w_1 + 2\,u)\,(w_1 + 2\,w)\,(x + y)}
                                      {(u + w)\,(w_1 - 2\,x)\,(w_1 - 2\,y)}\biggr]\biggr\}
                \,\exp[-r\,(u + w + x + y)]\,,\nonumber \\ [2ex]
         F_4 &=& {\rm Ei}[-2\,r\,(u + w)]\,\exp[r\,(u + w - x - y)]
                 +\biggl\{{\rm Ei}[-2\,r\,(x + y)]-{\rm Ei}[-r\,(w_1 + 2\,x)]\nonumber \\
              & &-{\rm Ei}[-r\,(w_1 + 2\,y)]
                 -\ln\biggl[\frac{(w_1 + 2\,u)\,(w_1 + 2\,w)\,(x + y)}
                                       {(u + w)\,(w_1 + 2\,x)\,(w_1 + 2\,y)}\biggr]\biggr\}
                \,\exp[-r\,(u + w - x - y)]\,,\nonumber \\ \label{07}
\end{eqnarray}
and ${\rm Ei}$ is the exponential integral function. The function $f(r)$ is the solution of
this differential equation, which vanishes at the small $r$ and has a Taylor expansion in $w_1$.

From the differential equation (\ref{04}) and similar differential equations
for derivatives of $f$ over nonlinear parameters, one can obtain recurrence relations in $n_i$
for the integrals $f(r,n_0,n_1,n_2,n_3,n_4))$ \cite{rec_h2, rec2_h2}.
These recurrences however, have various spurious singularities which make their practical use very cumbersome.
For these reason in our former calculations we have generated explicit expression for
derivatives of the master integral for three special cases.
The James-Coolidge (JC) basis, where $x=y=0$ \cite{bo_h2};
the generalized Heitler-London basis, where $x = w, y = -u$ \cite{bo_h2};
and the HeH$^+$ basis, where $x = w, y = u$ \cite{bo_hehp}, and in all these cases $w_1 = 0$ is assumed.
The results obtained for the ground electronic states of H$_2$ and HeH$^+$ were accurate
to $10^{-12} - 10^{-15}$ au.
These special cases do not work well for excited states in the intermediate region of $r$,
where the general basis with arbitrary nonlinear parameters is needed.
In a more recent work \cite{exp_h2}, we have developed a computational method for
exponentially correlated basis functions, namely that with $w_1 \neq 0$.
This basis is very flexible and can be used for the calculations of relativistic and QED effects.
However, the high computational cost and spurious singularities make this basis
rather difficult in application.

\section{Taylor expansion approach}
Here we overcame the above problems, and were able to present an efficient way to calculate two-centre two-electron integrals
with Ko\l os-Wolniewicz functions (\ref{01}) by using the Taylor expansion in $r$. This Taylor
expansion has already been proposed in Ref. \cite{rec2_h2}, and here we prove that it works in practice
by the calculation of Born-Oppenheimer energies for excited states of H$_2$.
First of all, this expansion is absolutely convergent
for an arbitrary $r$ \cite{fh}. In the typical situation near the equilibrium distance
$r = 1.4$ au one needs about 60 terms to obtain the integral with quadruple precision.
For larger distances ($r\sim 12$ au) the number of terms grows to about 200.
Coefficients of the expansion are obtained using arbitrary precision arithmetic \cite{mpfun}.
The evaluation time of all integrals on a single Intel Xenon core
for the largest basis of 3003 functions, using 64 digits arithmetic, was about 30 minuts, in comparison to
about 60 minutes of quadruple precision linear algebra (DSPGVX, LAPACK routine \cite{lapack}).
In fact the main issue in these calculations is not the evaluation time of integrals,
but the numerical instabilities in decomposition of the overlap matrix.

The complete set of recursions for the Taylor expansion of $f(r)$ and its derivatives
with respect to parameters can be obtained from the above differential equation.
It is, however, more convenient to use the following formulas for derivatives which were obtained in Ref. \cite{exp_h2}
\begin{eqnarray}\label{08}
(w_1^2-4\,w^2)\,\frac{\partial f'(r)}{\partial w} &=&
-\frac{r\,f(r)}{2}\,\frac{\partial\sigma_{02}}{\partial w}
-2\,r\,w\,f''(r) + \frac{-F_1 - F_2 + F_3 + F_4}{2}\,,\nonumber \\
(w_1^2-4\,u^2)\,\frac{\partial f'(r)}{\partial u} &=&
-\frac{r\,f(r)}{2}\,\frac{\partial\sigma_{02}}{\partial u}
-2\,r\,u\,f''(r) + \frac{F_1 + F_2 + F_3 + F_4}{2}\,,\nonumber \\
(w_1^2-4\,x^2)\,\frac{\partial f'(r)}{\partial x} &=&
-\frac{r\,f(r)}{2}\,\frac{\partial\sigma_{02}}{\partial x}
-2\,r\,x\,f''(r) + \frac{F_1 - F_2 + F_3 - F_4}{2}\,,\nonumber \\
(w_1^2-4\,y^2)\,\frac{\partial f'(r)}{\partial y} &=&
-\frac{r\,f(r)}{2}\,\frac{\partial\sigma_{02}}{\partial y}
-2\,r\,y\,f''(r) + \frac{-F_1 + F_2 + F_3 - F_4}{2}\,,
\end{eqnarray}
where $\sigma_{02}$ is defined in Eq. (\ref{05}),
the inhomogeneous terms $F_i$ terms are given by Eq.~(\ref{07}) and
$f'(r)$ means derivative with respect to $r$.
The derivative with respect to $w_1$ can
be easily obtained from the fact that function $r^{-2}\,f(r)$ is dimensionless, which entails that
\begin{equation}\label{09}
w_1\,\frac{\partial f(r)}{\partial w_1} = -w\,\frac{\partial f(r)}{\partial w}
-u\,\frac{\partial f(r)}{\partial u} -x\,\frac{\partial f(r)}{\partial x}
-y\,\frac{\partial f(r)}{\partial y} + r\,f'(r) - 2\,f(r)\,.
\end{equation}

The recurrence relations for the Taylor expansion of $f(r,n_0,n_1,n_2,n_3,n_4)$ are obtained as follows.
The master integral $f(r)$ has the following expansion in $r$:
\begin{equation}
f(r) = \sum_{n=1}^\infty \bigl[f^{(1)}_n\,(\ln (r) +\gamma_{\rm E}) + f^{(2)}_n\bigr]\,r^n\,, \label{10}
\end{equation}
(where $\gamma_{\rm E}$ is the Euler constant),
and so have all its derivatives $f(r,n_0,n_1,n_2,n_3,n_4)$.
To get recursions in $n_0$ we divide each equation in (\ref{08}) by the corresponding parameter,
sum up and use Eq. (\ref{09}). The resulting equation expanded in $r$ is
\begin{eqnarray}
f^{(1)}_n(n_0, 0, 0, 0, 0) &=& \frac{1}{4\,n\,(n+n_0)}\,
\biggl(F^{(1)}_{n-1}(n_0) - 4\,\frac{\sigma_{00}}{\sigma_{20}}\,f^{(1)}_{n-2}(n_0, 0, 0, 0, 0) \nonumber \\ &&
       + \frac{n}{u}\,f^{(1)}_n(n_0-2, 0, 0, 1, 0)
       + \frac{n}{w}\,f^{(1)}_n(n_0-2, 0, 0, 0, 1)\nonumber \\ &&
       + \frac{n}{x}\,f^{(1)}_n(n_0-2, 0, 1, 0, 0)
       + \frac{n}{y}\,f^{(1)}_n(n_0-2, 1, 0, 0, 0)\biggr)\,,\label{11}\\
f^{(2)}_n(n_0, 0, 0, 0, 0) &=& \frac{1}{4\,n\,(n+n_0)}\,
\biggl(F^{(2)}_{n-1}(n_0) - \frac{4\,\sigma_{00}}{\sigma_{20}}\,f^{(2)}_{n-2}(n_0, 0, 0, 0, 0)  \nonumber \\ &&
 - 4\,(2\,n + n_0)\,f^{(1)}_n(n_0, 0, 0, 0, 0) \nonumber \\ &&
 + \frac{1}{u}\bigl(f^{(1)}_n(n_0-2, 0, 0, 1, 0) + n\,f^{(2)}_n(n_0-2, 0, 0, 1, 0)\bigr) \nonumber \\ &&
 + \frac{1}{w}\bigl(f^{(1)}_n(n_0-2, 0, 0, 0, 1) + n\,f^{(2)}_n(n_0-2, 0, 0, 0, 1)\bigr) \nonumber \\ &&
 + \frac{1}{x}\bigl(f^{(1)}_n(n_0-2, 0, 1, 0, 0) + n\,f^{(2)}_n(n_0-2, 0, 1, 0, 0)\bigr) \nonumber \\ &&
 + \frac{1}{y}\bigl(f^{(1)}_n(n_0-2, 1, 0, 0, 0) + n\,f^{(2)}_n(n_0-2, 1, 0, 0, 0)\bigr)\biggr)\,, \label{12}
\end{eqnarray}
where
\begin{eqnarray}
F^{(i)}_{n}(n_0) &=& \frac{1}{2}\,\biggl[
  \biggl(\frac{1}{u} - \frac{1}{w} + \frac{1}{x} - \frac{1}{y}\biggr)\,F_{1,n}^{(i)}(n_0,0,0,0,0)
+ \biggl(\frac{1}{u} - \frac{1}{w} - \frac{1}{x} + \frac{1}{y}\biggr)\,F_{2,n}^{(i)}(n_0,0,0,0,0) \nonumber \\ &&
+ \biggl(\frac{1}{u} + \frac{1}{w} + \frac{1}{x} + \frac{1}{y}\biggr)\,F_{3,n}^{(i)}(n_0,0,0,0,0)
+ \biggl(\frac{1}{u} + \frac{1}{w} - \frac{1}{x} - \frac{1}{y}\biggr)\,F_{4,n}^{(i)}(n_0,0,0,0,0)\biggr]\,.
\nonumber \\ \label{13}
\end{eqnarray}

Recursions in $n_1, n_2, n_3$ and $n_4$ are obtained by differentiation of the corresponding equation.
For example, in $n_4$ they are the following
\begin{align}
&f^{(1)}_n(n_0, n_1, n_2, n_3, n_4+1) = -\frac{1}{4\,n\,(n_4+1)\,w^2}\,\biggl[\nonumber \\&
     2\,f^{(1)}_{n-2}(n_0, n_1-2, n_2, n_3, n_4-1) - 2\,w\,f^{(1)}_{n-2}(n_0, n_1-2, n_2, n_3, n_4)\nonumber \\&
     -4\,f^{(1)}_{n-2}(n_0, n_1-1, n_2-1, n_3-1, n_4) + 4\,u\,f^{(1)}_{n-2}(n_0, n_1-1, n_2-1, n_3, n_4)\nonumber \\&
     +4\,x\,f^{(1)}_{n-2}(n_0, n_1-1, n_2, n_3-1, n_4) - 4\,y\,f^{(1)}_{n-2}(n_0, n_1-1, n_2, n_3, n_4-1)\nonumber \\&
     -4\,(u\,x - w\,y)\,f^{(1)}_{n-2}(n_0, n_1-1, n_2, n_3, n_4) + 2\,f^{(1)}_{n-2}(n_0, n_1, n_2-2, n_3, n_4-1)\nonumber \\&
     -2\,w\,f^{(1)}_{n-2}(n_0, n_1, n_2-2, n_3, n_4) + 4\,y\,f^{(1)}_{n-2}(n_0, n_1, n_2-1, n_3-1, n_4)\nonumber \\&
     -4\,x\,f^{(1)}_{n-2}(n_0, n_1, n_2-1, n_3, n_4-1) - 4\,(u\,y - x\,w)\,f^{(1)}_{n-2}(n_0, n_1, n_2-1, n_3, n_4)\nonumber \\&
     +2\,f^{(1)}_{n-2}(n_0, n_1, n_2, n_3-2, n_4-1) - 2\,w\,f^{(1)}_{n-2}(n_0, n_1, n_2, n_3-2, n_4)\nonumber \\&
     -4\,u\,f^{(1)}_{n-2}(n_0, n_1, n_2, n_3-1, n_4-1) + 4\,(u\,w - x\,y)\,f^{(1)}_{n-2}(n_0, n_1, n_2, n_3-1, n_4)\nonumber \\&
     -2\,f^{(1)}_{n-2}(n_0, n_1, n_2, n_3, n_4-3) + 6\,w\,f^{(1)}_{n-2}(n_0, n_1, n_2, n_3, n_4-2)\nonumber \\&
     +2\,(u^2 - 3\,w^2 + x^2 + y^2)\,f^{(1)}_{n-2}(n_0, n_1, n_2, n_3, n_4-1) \nonumber \\&
     +2\,(w^3 - u^2\,w - w\,x^2 + 2\,u\,x\,y - w\,y^2)\,f^{(1)}_{n-2}(n_0, n_1, n_2, n_3, n_4) \nonumber \\&
     -n\,(n_4+1)\,f^{(1)}_{n}(n_0-2, n_1, n_2, n_3, n_4+1) - 2\,n\,(n - 2\,n_4+1)\,f^{(1)}_{n}(n_0, n_1, n_2, n_3, n_4-1)\nonumber \\&
     + 2\,n\,(n - 4\,n_4 - 1)\,w\,f^{(1)}_{n}(n_0, n_1, n_2, n_3, n_4) -\frac{1}{2}\bigl[-F^{(1)}_{1,n-1}(n_0, n_1, n_2, n_3, n_4)\nonumber \\&
     - F^{(1)}_{2,n-1}(n_0, n_1, n_2, n_3, n_4) + F^{(1)}_{3,n-1}(n_0, n_1, n_2, n_3, n_4) + F^{(1)}_{4,n-1}(n_0, n_1, n_2, n_3, n_4)\bigr]
    \biggr]\,, \label{14}
\end{align}
\begin{align}
&f^{(2)}_n(n_0, n_1, n_2, n_3, n_4+1) = -\frac{1}{4\,n\,(n_4+1)\,w^2}\,\biggl[\nonumber \\&
     -(n_4+1)\,f^{(1)}_{n}(n_0-2, n_1, n_2, n_3, n_4+1) - 2\,(2\,n - 2\,n_4 + 1)\,f^{(1)}_{n}(n_0, n_1, n_2, n_3, n_4-1)\nonumber \\ &
     + 2\,(2\,n - 4\,n_4 - 1)\,w\,f^{(1)}_{n}(n_0, n_1, n_2, n_3, n_4) + 4\,(n_4+1)\,w^2\,f^{(1)}_{n}(n_0, n_1, n_2, n_3, n_4+1)\nonumber \\ &
     + 2\,f^{(2)}_{n-2}(n_0, n_1-2, n_2, n_3, n_4-1) - 2\,w\,f^{(2)}_{n-2}(n_0, n_1-2, n_2, n_3, n_4)\nonumber \\ &
     - 4\,f^{(2)}_{n-2}(n_0, n_1 - 1, n_2 - 1, n_3 - 1, n_4) + 4\,u\,f^{(2)}_{n-2}(n_0, n_1 - 1, n_2 - 1, n_3, n_4)\nonumber \\ &
     + 4\,x\,f^{(2)}_{n-2}(n_0, n_1 - 1, n_2, n_3 - 1, n_4) - 4\,y\,f^{(2)}_{n-2}(n_0, n_1 - 1, n_2, n_3, n_4 - 1)\nonumber \\ &
     - 4\,(u\,x - w\,y)\,f^{(2)}_{n-2}(n_0, n_1 - 1, n_2, n_3, n_4) + 2\,f^{(2)}_{n-2}(n_0, n_1, n_2- 2, n_3, n_4 - 1)\nonumber \\ &
     - 2\,w\,f^{(2)}_{n-2}(n_0, n_1, n_2 - 2, n_3, n_4) + 4\,y\,f^{(2)}_{n-2}(n_0, n_1, n_2 - 1, n_3 - 1, n_4)\nonumber \\ &
     - 4\,x\,f^{(2)}_{n-2}(n_0, n_1, n_2- 1, n_3, n_4 - 1) - 4\,(u\,y - x\,w)\,f^{(2)}_{n-2}(n_0, n_1, n_2 - 1, n_3, n_4)\nonumber \\ &
     + 2\,f^{(2)}_{n-2}(n_0, n_1, n_2, n_3 - 2, n_4 - 1) - 2\,w\,f^{(2)}_{n-2}(n_0, n_1, n_2, n_3 - 2, n_4)\nonumber \\ &
     - 4\,u\,f^{(2)}_{n-2}(n_0, n_1, n_2, n_3 - 1, n_4 - 1) + 4\,(u\,w - x\,y)\,f^{(2)}_{n-2}(n_0, n_1, n_2, n_3 - 1, n_4)\nonumber \\ &
     - 2\,f^{(2)}_{n-2}(n_0, n_1, n_2, n_3, n_4 - 3) + 6\,w\,f^{(2)}_{n-2}(n_0, n_1, n_2, n_3, n_4 - 2) \nonumber \\ &
     + 2\,(u^2 - 3\,w^2 + x^2 + y^2)\,f^{(2)}_{n-2}(n_0, n_1, n_2, n_3, n_4 - 1)\nonumber \\ &
     + 2\,(w^3 - u^2\,w - w\,x^2 + 2\,u\,x\,y - w\,y^2)\,f^{(2)}_{n-2}(n_0, n_1, n_2, n_3, n_4)\nonumber \\ &
     - n\,(n_4 + 1)\,f^{(2)}_{n}(n_0 - 2, n_1, n_2, n_3, n_4 + 1)
     - 2\,n\,(n - 2\,n_4 + 1)\,f^{(2)}_{n}(n_0, n_1, n_2, n_3, n_4 - 1)\nonumber \\ &
     + 2\,n\,(n - 4\,n_4 - 1)\,w\,f^{(2)}_{n}(n_0, n_1, n_2, n_3, n_4)
     - \frac{1}{2}\bigl[-F^{(2)}_{1,n-1}(n_0, n_1, n_2, n_3, n_4)\nonumber \\&
     - F^{(2)}_{2,n-1}(n_0, n_1, n_2, n_3, n_4) + F^{(2)}_{3,n-1}(n_0, n_1, n_2, n_3, n_4) + F^{(2)}_{4,n-1}(n_0, n_1, n_2, n_3, n_4)\bigr]
    \biggr]\,, \label{15}
\end{align}
where the inhomogeneous terms $F^{(i)}_{k,n}(n_0, n_1, n_2, n_3, n_4)$ are constructed in Appendix A.
Together with recursion relations in $n_1$, $n_2$, and $n_3$ they allow for the calculation
of all integrals with the condition that $x$ and $y$ are not close to 0.
In practice we assume that both $x$ and $y$ are greater than $0.01$.
The case where $x=0$ or $y=0$ is considered in the next Section.

\section{Special cases}
The obtained recursions work in general cases, with some exceptions.
When one of the parameters is close to zero, then these recursions become unstable.
We have not been able to find recursions which are safe at small $x$ and $y$
and its evaluation is linear in the length of the Taylor series.
Instead, one can perform a Taylor expansion in a small parameter, for example in $x$.
Coefficients can be obtained from the master differential equation (\ref{04})
by differentiating over $w_1$ and $x$ at $w_1 = x=0$. It becomes then an algebraic equation
for $f(n_0,0, n_2,0,0)$. For example the result for the master integral is
\begin{eqnarray}
f(r) &=& \frac{\sinh(r\,y)}{r\,y}\,\frac{1}{4\,u\,w}\,
\biggl\{-{\rm Ei}(-2\,r\,u)\,\exp(r\,(u - w)) -
      {\rm Ei}(-2\,r\,w)\,\exp(r\,(w-u)) \nonumber \\ && +
      {\rm Ei}(-2\,r\,(u + w))\,\exp(r\,(u + w)) +
      \biggl[\ln\biggl(\frac{2\,r\,u\,w}{u + w}\biggr)+\gamma_{\rm E}\biggr]\,\exp(-r\,(u + w))\biggr\}.
\label{18}
\end{eqnarray}
This expression can also be derived from the Neumann expansion of $1/r_{12}$ in spherical oblate
coordinates, see also Appendix B, in which the master integral $f$
is presented in terms of the Neumann expansion. In the calculation performed here,
we do not derive explicit expressions for integrals with  $w_1 = x=0$, but
adopt the numerical recursion for the Taylor series for the case of $x=0$.
This is done by differentiating the third equation in (\ref{08}) with respect to
all parameters at $w_1 = x = 0$. It then becomes an equation
for $f^{(i)}_n(n_0,n_1,n_2,n_3,n_4)$, which gives recursions at $x=0$, namely
\begin{align}
&f^{(1)}_n(n_0, n_1, n_2, n_3, n_4) = \frac{1}{4\,u\,w\,y}\,\biggl[
- 2\,f^{(1)}_n(n_0, n_1-2, n_2-1, n_3, n_4) \nonumber \\ &
+ 4\,y\,f^{(1)}_n(n_0, n_1-1, n_2-1, n_3, n_4)
+ 4\,f^{(1)}_n(n_0, n_1-1, n_2, n_3-1, n_4-1) \nonumber \\ &
- 4\,w\,f^{(1)}_n(n_0, n_1-1, n_2, n_3-1, n_4)
- 4\,u\,f^{(1)}_n(n_0, n_1-1, n_2, n_3, n_4-1) \nonumber \\ &
+ 4\,u\,w\,f^{(1)}_n(n_0, n_1-1, n_2, n_3, n_4)
+ 2\,f^{(1)}_n(n_0, n_1, n_2-3, n_3, n_4) \nonumber \\ &
- 2\,f^{(1)}_n(n_0, n_1, n_2-1, n_3-2, n_4)
+ 4\,u\,f^{(1)}_n(n_0, n_1, n_2-1, n_3-1, n_4) \nonumber \\ &
- 2\,f^{(1)}_n(n_0, n_1, n_2-1, n_3, n_4-2)
+ 4\,w\,f^{(1)}_n(n_0, n_1, n_2-1, n_3, n_4-1) \nonumber \\ &
- 2\,(u^2 + w^2 + y^2)\,f^{(1)}_n(n_0, n_1, n_2-1, n_3, n_4)
- 4\,y\,f^{(1)}_n(n_0, n_1, n_2, n_3-1, n_4-1) \nonumber \\ &
+ 4\,w\,y\,f^{(1)}_n(n_0, n_1, n_2, n_3-1, n_4)
+ 4\,u\,y\,f^{(1)}_n(n_0, n_1, n_2, n_3, n_4-1) \nonumber \\ &
+ (n+2)\,(n_2+1)\,f^{(1)}_{n+2}(n_0-2, n_1, n_2+1, n_3, n_4) \nonumber \\ &
+ 2\,(n+2)\,(n - 2\,n_2+3)\,f^{(1)}_{n+2}(n_0, n_1, n_2-1, n_3, n_4)
+ \frac{1}{2}\bigl[
 F^{(1)}_{1,n+1}(n_0, n_1, n_2, n_3, n_4)\nonumber \\ &
-F^{(1)}_{2,n+1}(n_0, n_1, n_2, n_3, n_4)
+F^{(1)}_{3,n+1}(n_0, n_1, n_2, n_3, n_4)
-F^{(1)}_{4,n+1}(n_0, n_1, n_2, n_3, n_4)
\bigr]\biggr]\,,
\end{align}
\begin{align}
&f^{(2)}_n(n_0, n_1, n_2, n_3, n_4) = \frac{1}{4\,u\,w\,y}\,\biggl[
  (n_2+1)\,f^{(1)}_{n+2}(n_0-2, n_1, n_2+1, n_3, n_4) \nonumber \\ &
+ 2\,(2\,(n+2) - 2\,n_2+1)\,f^{(1)}_{n+2}(n_0, n_1, n_2-1, n_3, n_4)
- 2\,f^{(2)}_n(n_0, n_1-2, n_2-1, n_3, n_4) \nonumber \\ &
+ 4\,y\,f^{(2)}_n(n_0, n_1-1, n_2-1, n_3, n_4)
+ 4\,f^{(2)}_n(n_0, n_1-1, n_2, n_3-1, n_4-1) \nonumber \\ &
- 4\,w\,f^{(2)}_n(n_0, n_1-1, n_2, n_3-1, n_4)
- 4\,u\,f^{(2)}_n(n_0, n_1-1, n_2, n_3, n_4-1) \nonumber \\ &
+ 4\,u\,w\,f^{(2)}_n(n_0, n_1-1, n_2, n_3, n_4)
+ 2\,f^{(2)}_n(n_0, n_1, n_2-3, n_3, n_4)  \nonumber \\ &
- 2\,f^{(2)}_n(n_0, n_1, n_2-1, n_3-2, n_4)
+ 4\,u\,f^{(2)}_n(n_0, n_1, n_2-1, n_3-1, n_4)  \nonumber \\ &
- 2\,f^{(2)}_n(n_0, n_1, n_2-1, n_3, n_4-2)
+ 4\,w\,f^{(2)}_n(n_0, n_1, n_2-1, n_3, n_4-1)  \nonumber \\ &
- 2\,(u^2 + w^2 + y^2)\,f^{(2)}_n(n_0, n_1, n_2-1, n_3, n_4)
- 4\,y\,f^{(2)}_n(n_0, n_1, n_2, n_3-1, n_4-1)  \nonumber \\ &
+ 4\,w\,y\,f^{(2)}_n(n_0, n_1, n_2, n_3-1, n_4)
+ 4\,u\,y\,f^{(2)}_n(n_0, n_1, n_2, n_3, n_4-1)  \nonumber \\ &
+ (n+2)\,(n_2+1)\,f^{(2)}_{n+2}(n_0-2, n_1, n_2+1, n_3, n_4) \nonumber \\ &
+ 2\,(n+2)\,(n - 2\,n_2+3)\,f^{(2)}_{n+2}(n_0, n_1, n_2-1, n_3, n_4)
+ \frac{1}{2}\bigl[
 F^{(2)}_{1,n+1}(n_0, n_1, n_2, n_3, n_4)\nonumber \\ &
-F^{(2)}_{2,n+1}(n_0, n_1, n_2, n_3, n_4)
+F^{(2)}_{3,n+1}(n_0, n_1, n_2, n_3, n_4)
-F^{(2)}_{4,n+1}(n_0, n_1, n_2, n_3, n_4)
\bigr]\biggr]\,.
\end{align}

The assumption of a symmetry with respect to $A\leftrightarrow B$ introduces integrals with
$x=y=0$, and these have already been derived using recursion relations obtained in Ref. \cite{bo_h2}.
Moreover, functions with $x=y=0$, the so-called JC basis, work very well
for short internuclear distances ($r<6$) au, as is demonstrated in the next Section.
They can also be obtained via the Taylor series, but most probably
the best way to calculate integrals with small $x$ or $y$ is
by the Neumann expansion, which becomes finite at  $x=0$ or $y=0$.

For very large internuclear distances, $x$ and $y$ are significantly different from 0
and the Taylor expansion requires many terms.
We tried to use a generalized Heitler-London basis, where $x=\pm w$ and $y=\pm u$. The analytic expression
for this special type of integral was obtained using recursion relations derived in Ref. \cite{rec_h2}.
The expression for $f(n_0,n_1,n_2,n_3,n_4)$ involves exponential integral function Ei,
the exponential and the master integral $f(r)$. In spite of using explicit expressions,
this way of calculation is not much more effective than the Taylor series. This is because
the analytic expressions are very long for large $n_0$
and their evaluation requires the higher precision arithmetic.
One cannot exclude that there is a clever way to express long polynomials in terms of
some known functions, as is the case for odd $n_0$ \cite{rec2_h2},
but so far we have not been able to do so.
In the appendix B we present a compact Neumann representation for the master integral,
which can be helpful in developing an approach without the use of the Taylor expansion.

 \section{Numerical results}
The wave function of electronic $\Sigma$ state of the H$_2$ molecule is expressed in
terms of the basis wave functions $\phi_i$ in Eq. (\ref{01}) as follows
\begin{equation}
\psi_\Sigma = \sum_i c_i\, (1\pm P_{AB})\,(1\pm P_{12})\,\phi_i\,, \label{19}
\end{equation}
where $P_{AB}$ permutes the nuclei $A$ and $B$, $P_{12}$ interchanges the two electrons,
and $c_i$ are linear coefficients, obtained as components of the eigenvector of the Hamiltonian matrix.
The powers $n_i$ of electronic distances in $\phi$ are chosen by the condition
\begin{equation}
\sum_{i=0}^4 n_i \leq \Omega\,, \label{20}
\end{equation}
with the parameter $\Omega$ changing from 5 to 10. The largest value $\Omega = 10$ corresponds
to the 3003 length of the general basis and to 1910 in the JC basis.
The JC basis is more compact, because the symmetry $P_{AB}$ of $\Sigma^+_g$ state
restricts certain combinations of $n_i$, which is not the case of the basis with $x$ or $y \neq 0$.

\begin{table}
\caption{Nonrelativistic Born-Oppenheimer energy $E(r)$ of excited electronic $^1\Sigma_g^+$ states of the H$_2$ molecule
at four internuclear distances $r$, in a.u., calculated with $\Omega=9,10$.
The results are compared with the previous best available data obtained with
KW \cite{OSW99, sigma} and ECG \cite{KC03} wave functions.
\label{table1}}
\begin{ruledtabular}
  \begin{tabular}{cc....}
\multicolumn{1}{c}{State}  & \multicolumn{1}{c}{$\Omega$}
& \multicolumn{1}{c}{$E(1.5)$} & \multicolumn{1}{c}{$E(3.0)$}
& \multicolumn{1}{c}{$E(6.0)$} & \multicolumn{1}{c}{$E(12.0)$}\\ \hline
$EF\,^1\Sigma_g^+$       &9&  -0.703\,x000\,246\,8   & -0.690\,x747\,055\,9  & -0.694\,x267\,016  & -0.628\,x742\,038 \\
                        &10& -0.703\,x000\,247\,0   & -0.690\,x747\,056\,3  & -0.694\,x267\,029  & -0.628\,x742\,088 \\
                        &ECG & -0.703\,x000\,229      & -0.690\,x747\,014     &                   -0.694\,x267\,005 & -0.628\,x742\,051                   \\
                        &KW& -0.702\,x999\,909      & -0.690\,x746\,981     &-0.694\,x263\,365  & -0.628\,x730\,759  \\[1ex]
$GK\,^1\Sigma_g^+$       &9&  -0.639\,x008\,658\,8   & -0.656\,x985\,931     & -0.626\,x147\,966  & -0.624\,x745\,652 \\
                        &10& -0.639\,x008\,659\,5   & -0.656\,x985\,945     & -0.626\,x147\,969  & -0.624\,x745\,653 \\
                        &KW& -0.639\,x007\,737      & -0.656\,x983\,847     & -0.626\,x147\,852  &                   \\[1ex]
$H\bar H\,^1\Sigma_g^+$  &9&  -0.636\,x334\,418\,0   & -0.630\,x554\,122     & -0.583\,x463\,524  & -0.604\,x584\,075 \\
                        &10& -0.636\,x334\,418\,5   & -0.630\,x554\,134     & -0.583\,x463\,574  & -0.604\,x584\,280 \\
                        &KW& -0.636\,x333\,766      & -0.630\,x550\,821     & -0.583\,x461\,341  & -0.604\,x529\,322 \\[1ex]
$P\,^1\Sigma_g^+$        &9&  -0.614\,x073\,200      & -0.623\,x922\,685     & -0.564\,x424\,571  & -0.555\,x682\,254 \\
                        &10& -0.614\,x073\,206      & -0.623\,x922\,735     & -0.564\,x424\,574  & -0.555\,x682\,256 \\
                        &KW& -0.614\,x049\,795      & -0.623\,x917\,301     & -0.564\,x423\,608  &                   \\[1ex]
$O\,^1\Sigma_g^+$        &9&  -0.612\,x886\,933      & -0.607\,x984\,433     & -0.553\,x905\,260  & -0.555\,x533\,720 \\
                        &10& -0.612\,x886\,935      & -0.607\,x984\,513     & -0.553\,x905\,272  & -0.555\,x533\,721 \\
                        &KW& -0.612\,x885\,514      & -0.607\,x841\,139     & -0.553\,x862\,823  &                   \\
\end{tabular}
\end{ruledtabular}
\end{table}

Table \ref{table1} presents numerical results for the nonrelativistic energy
of the first 5 excited electronic $^1\Sigma_g^+$ states
of the H$_2$ molecule obtained for $r=1.5, 3.0$ au with the JC, and for $r=6,12$ au with the general basis.
The nonlinear parameters have been optimized against the binding energies for $\Omega = 7$.
In the case of the JC basis, we used a double basis sets with two independent set of parameters,
and $\Omega_2 = \Omega_1-2$.
Since the minimization leads to $u\approx w$ in the second set, we impose the additional condition
for $n_i$ to eliminate linear dependence of the basis for $u=w$.
In the calculations involving general basis, we used only one set of parameters.

In spite of the use of a very simple basis with just 4 independent nonlinear parameters,
our results are two- or three-orders of magnitude more accurate than
the most accurate results obtained so far, namely those in the Ko\l os-Wolniewicz basis \cite{OSW99, sigma} and
explicitly correlated Gaussian (ECG) functions \cite{KC03}.
This improved accuracy is especially visible for higher excited states, see Table \ref{table1}.

\section{Summary}

In this paper we have developed a numerical procedure for two-centre two-electron integrals
with exponential functions. It is based on the Taylor expansion in the internuclear distance,
differential equations for derivatives of the master integral
with respect to nonlinear parameters (\ref{08}), and recursion relations.
The whole code is very compact but requires a high precision arithmetic, typically 64 digits.
This numerical approach has been applied to the calculation of excited $^1\Sigma^+_g$ states
of the H$_2$ molecule. The obtained results, in a relatively small and simple basis, are more accurate
than any best previous results, see Table \ref{table1}.

As well as being simple, this numerical approach can be extended, we think,
to integrals with additional inverse powers of electronic distances
which are needed for the calculations of relativistic and QED corrections, including the yet unknown
$\alpha^4\,$Ry contribution which limits the present accuracy of theoretical predictions for H$_2$.

\appendix
\section{Taylor expansion of the inhomogeneous terms}
The general recursion relations for the Taylor expansion of two-centre two-electron integrals
involve inhomogeneous terms. They are split into the logarithmic
and nonlogarithmic parts similarly to Eq. (\ref{10}).
The logarithmic part with $\sigma^{(1)} = \delta_{n_0}\,[1 - (-1)^n]/(n_1!\,n_2!\,n_3!\,n_4!)$ is
\begin{eqnarray}
F^{(1)}_{1,n}(n_0, n_1, n_2, n_3, n_4) &=&
\sigma^{(1)}\,(-1)^{n_2 + n_3}\,\frac{(u - w + x - y)^{n - n_1 - n_2 - n_3 - n_4}}{(n - n_1 - n_2 - n_3 - n_4)!}\,,
 \label{A1} \\ \nonumber \\
F^{(1)}_{2,n}(n_0, n_1, n_2, n_3, n_4) &=&
\sigma^{(1)}\,(-1)^{n_1 + n_3}\,\frac{(u - w - x + y)^{n - n_1 - n_2 - n_3 - n_4}}{(n - n_1 - n_2 - n_3 - n_4)!}\,,
\label{A2} \\ \nonumber \\
F^{(1)}_{3,n}(n_0, n_1, n_2, n_3, n_4) &=&
\sigma^{(1)}\,(-1)^{n_1 + n_2 + n_3 + n_4}\,\frac{(u + w + x + y)^{n - n_1 - n_2 - n_3 - n_4}}{(n - n_1 - n_2 - n_3 - n_4)!}\,,
 \label{A3} \\ \nonumber \\
F^{(1)}_{4,n}(n_0, n_1, n_2, n_3, n_4) &=&
\sigma^{(1)}\,(-1)^{n_3 + n_4}\,\frac{(u + w - x - y)^{n - n_1 - n_2 - n_3 - n_4}}{(n - n_1 - n_2 - n_3 - n_4)!}\,,
\label{A4} \\ \nonumber
\end{eqnarray}
and the nonlogarithmic part with $\sigma^{(2)} = 1/(n_0!\,n_1!\,n_2!\,n_3!\,n_4!)$ is
\begin{eqnarray}
F^{(2)}_{1,n}(n_0, n_1, n_2, n_3, n_4) &=& \sigma^{(2)}\,\bigl[
 (-1)^{n_2}\,F_E(-u, -w + x - y, n_0, n_3, n - n_1 - n_2 - n_4) \nonumber \\ &&
- (-1)^{n_1}\,F_E(-w, -u - x + y, n_0, n_4, n - n_1 - n_2 - n_3) \nonumber \\ &&
+ (-1)^{n_2}\,F_L(-u, -w + x - y, n_0, n_3, n - n_1 - n_2 - n_4) \nonumber \\ &&
- (-1)^{n_1}\,F_L(-w, -u - x + y, n_0, n_4, n - n_1 - n_2 - n_3)\bigr]\,, \label{A5}
\end{eqnarray}
\begin{eqnarray}
F^{(2)}_{2,n}(n_0, n_1, n_2, n_3, n_4) &=& \sigma^{(2)}\,\bigl[
(-1)^{n_1}\,F_E(-u, -w - x + y, n_0, n_3, n - n_1 - n_2 - n_4) \nonumber \\ &&
- (-1)^{n_2}\,F_E(-w, -u + x - y, n_0, n_4, n - n_1 - n_2 - n_3) \nonumber \\ &&
+ (-1)^{n_1}\,F_L(-u, -w - x + y, n_0, n_3, n - n_1 - n_2 - n_4) \nonumber \\ &&
- (-1)^{n_2}\,F_L(-w, -u + x - y, n_0, n_4, n - n_1 - n_2 - n_3)\bigr]\,, \label{A6}
\end{eqnarray}
\begin{eqnarray}
F^{(2)}_{3,n}(n_0, n_1, n_2, n_3, n_4) &=& \sigma^{(2)}\,\bigl[
(-1)^{n_1 + n_2}\,F_E(-u - w, x + y, 0, n_3 + n_4, n - n_1 - n_2)\,\delta_{n_0} \nonumber \\ &&
+ (-1)^{n_1 + n_2}\,F_E(x + y, -u - w, 0, n_1 + n_2, n - n_3 - n_4)\,\delta_{n_0} \nonumber \\ &&
+ (-1)^{n_1 + n_2}\,[1 + (-1)^n]\,F_L(-u - w, x + y, 0, n_3 + n_4, n - n_1 - n_2)\,\delta_{n_0} \nonumber \\ &&
- (-1)^{n_2}\,F_E(x, -u - w - y, n_0, n_2, n - n_1 - n_3 - n_4) \nonumber \\ &&
- (-1)^{n_1}\,F_E(y, -u - w - x, n_0, n_1, n - n_2 - n_3 - n_4) \nonumber \\ &&
- (-1)^{n + n_1 + n_2 + n_4}\,F_L(-u, w + x + y, n_0, n_3, n - n_1 - n_2 - n_4) \nonumber \\ &&
- (-1)^{n + n_1 + n_2 + n_3}\,F_L(-w, u + x + y, n_0, n_4, n - n_1 - n_2 - n_3)\bigr]\,, \label{A7}
\end{eqnarray}
\begin{eqnarray}
F^{(2)}_{4,n}(n_0, n_1, n_2, n_3, n_4) &=& \sigma^{(2)}\,\bigl[
F_E(-u - w, -x - y, 0, n_3 + n_4, n - n_1 - n_2)\,\delta_{n_0} \nonumber \\ &&
+ F_E(-x - y, -u - w, 0, n_1 + n_2, n - n_3 - n_4)\,\delta_{n_0} \nonumber \\ &&
+ [1 + (-1)^n]\,F_L(-u - w, -x - y, 0, n_3 + n_4, n - n_1 - n_2)\,\delta_{n_0} \nonumber \\ &&
- (-1)^{n_1}\,F_E(-x, -u - w + y, n_0, n_2, n - n_1 - n_3 - n_4) \nonumber \\ &&
- (-1)^{n_2}\,F_E(-y, -u - w + x, n_0, n_1, n - n_2 - n_3 - n_4) \nonumber \\ &&
- (-1)^{n + n_4}\,F_L(-u, w - x - y, n_0, n_3, n - n_1 - n_2 - n_4) \nonumber \\ &&
- (-1)^{n + n_3}\,F_L(-w, u - x - y, n_0, n_4, n - n_1 - n_2 - n_3)\bigr]\,, \label{A8}
\end{eqnarray}
where
\begin{eqnarray}
F_L(b,a;k,m,n) &=& \frac{\partial^k}{\partial c^k}\biggr|_{c=0}\,
                   \frac{\partial^m}{\partial b^m}\, \frac{(a-b)^n}{n!} \ln(-2 b - c)\,, \label{A9}\\
F_E(b,a;k,m,n) &=& \frac{\partial^k}{\partial c^k}\biggr|_{c=0}\,
                   \frac{\partial^m}{\partial b^m}\,
                   \frac{1}{n!}\frac{\partial^n}{\partial r^n}\biggr|_{r=1}\,
                   e^{r\,(a-b)}\, \Bigl({\rm Ei}\bigl[r(2\,b + c)\bigr] - \ln\bigl[r\,(-2 b - c)\bigr] - \gamma\Bigr)\,.
\nonumber\\ \label{A10}
\end{eqnarray}

\section{Neumann expansion of the master integral}
The Neumann expansion of $r_{12}^{-1}$ or $e^{-w_1\,r_{12}}\,r_{12}^{-1}$ in spherical oblate coordinates
has been previously used in the calculation of two-centre two-electron integrals,
see \cite{harris2} and references therein.
Here we present a compact formula for $f(r)$ at $w_1=0$
\begin{eqnarray}
f(r) &=& \sum_{n=0}^\infty \frac{(1+2\,n)}{4}\,r^2\,j_n(r\,x)\,j_n(r\,y)\,\Pi_n(r\,u,r\,w)\,, \label{B1}
\end{eqnarray}
where $j_n$, $h_n$ are modified spherical Bessel functions,
\begin{eqnarray}
j_n(x) &=& x^n\,\Bigl(\frac{1}{x}\,\frac{d}{d x}\Bigr)^n\,\frac{\sinh(x)}{x},\nonumber \\
h_n(x) &=& x^n\,\Bigl(\frac{1}{x}\,\frac{d}{d x}\Bigr)^n\,\frac{\exp(-x)}{x}\,, \label{17}
\end{eqnarray}
and where
\begin{eqnarray}
\Pi_{n}(u,w) &=& h_{n}(-u)\,h_{n}(-w)\,{\rm Ei}\bigl(-2\,(u+w)\bigr)
               + h_{n}(u)\,h_{n}(w)\,\biggl(\ln\Bigl(\frac{2\,u\,w}{u+w}\Bigr)+\gamma\biggr)
\nonumber \\ &&
              + (-1)^{n}\,h_{n}(-u)\,h_{n}(w)\,{\rm Ei}(-2\,u)
             + (-1)^{n}\,h_{n}(u)\,h_{n}(-w)\,{\rm Ei}(-2\,w)
\nonumber \\ &&
             + \frac{e^{-u-w}}{u\,w}\,W_{n}\Bigl(\frac{1}{u},\frac{1}{w}\Bigr)\,.
\label{B2}
\end{eqnarray}
$W_{n}(\alpha,\beta)$ is a polynomial in $\alpha$ and $\beta$, such that it eliminates $1/r$ singularity
in $\Pi_{n}$. It can be constructed recursively as follows,
\begin{equation}
W_{i}(\alpha,\beta) = -\sum_{k=1}^i \frac{2}{i-k+1}\,p_{k,i}(\alpha)\,p_{k,i}(\beta)\,, \label{B3}
\end{equation}
where $p_{k,i}$ are polynomials, for which the following recursions work
\begin{eqnarray}
p_{k,i}(x) &=& p^e_{k,i}(x)+p^o_{k,i}(x),\nonumber \\
p^e_{1,i}(x) &=& 1,\nonumber \\
p^o_{1,i}(x) &=& 0 ,\nonumber \\
p^e_{k+1,i}(x) &=& p^e_{k,i}(x) + \frac{1+(-1)^{k}}{2}\,x\,(2\,i-2\,k+1)\,p^o_{k,i}(x), \nonumber \\
p^o_{k+1,i}(x) &=& p^o_{k,i}(x) + \frac{1+(-1)^{k+1}}{2}\,x\,(2\,i-2\,k+1) \,p^e_{k,i}(x).
\label{B4}
\end{eqnarray}
We have not been able to prove that the above formula for $f(r)$ solves
the differential equation (\ref{B2}), instead  we have shown
that first terms of the Taylor expansion in $x$ coincides with that obtained
from the differential equation, see the text before Eq. (\ref{18}),
and we have checked its correctness numerically.
The principal advantage of this Neumann expansion
is the fact that differentiation with respect to nonlinear parameters
can be as easily performed as derivatives of spherical Bessel functions
and of the polynomial $W_n$, and this differentiation does not lead to any singularities.
In addition $\Pi_n(r)$ has very simple integral representation. Namely,
if we consider $\Pi_n(r)$ as an analytic function of $r$ with branch cut for $r<0$,
then it can  be expressed by the following dispersion relation
\begin{equation}
\Pi_{n}(r\,u, r\,w) =
\int_0^\infty dr'\,\left(\frac{1}{r} - \frac{1}{r+r'}\right)\,
4\,j_{n}(r'\,u)\,j_{n}(r'\,w)\,\exp\bigl(-(r+r')\,(u+w)\bigr)\,. \label{B5}
\end{equation}
which is convenient for the numerical evaluation.

\section*{Acknowledgments}
Author wishes to thank M. Lewin for the invitation
to the Institute of Henri Poincare, where this paper was written.
This work was supported by NCN grant 2012/04/A/ST2/00105.

\end{document}